\documentclass[sts]{imsart}

\RequirePackage{amsthm,amsmath,amsfonts,amssymb}
\RequirePackage[authoryear]{natbib}
\RequirePackage{graphicx}

\startlocaldefs
\theoremstyle{plain}

\theoremstyle{remark}

\DeclareMathOperator*{\argmax}{arg\,max}

\endlocaldefs

\begin{document}

\begin{frontmatter}
\title{A Statistical Test for Identifying Extreme Stationary Distribution Values in Markov Transition Matrices with Emphasis on Mobility Networks}
\runtitle{Identifying Extreme Stationary Distribution Values}

\begin{aug}
\author[A]{\fnms{Michael}~\snm{Kane}\ead[label=e1]{michael.kane@yale.edu}\orcid{http://orcid.org/0000-0003-1899-6662}},
\author[B]{\fnms{Owais}~\snm{Gilani}\ead[label=e2]{owais.gilani@bucknell.edu}\orcid{0000-0002-0402-6975}}
\and
\author[C]{\fnms{Simon}~\snm{Urbanek}\ead[label=e3]{s.urbanek@auckland.ac.nz}\orcid{0000-0003-2297-1732}}



\address[A]{Michael Kane is an Assistant Professor, Biostatistics Department, School of Public Health,  Yale University, New Haven, CT, USA\printead[presep={\ }]{e1}.}

\address[B]{Owais Gilani is an Associate Professor, Mathematics Department,
Bucknell University, Lewisburg, PA, USA\printead[presep={\ }]{e2}.}

\address[C]{Simon Urbanek is a Senior Lecturer and R Core Member, Statistics Department,
The University of Auckland, Auckland, NZ\printead[presep={\ }]{e3}.}

\end{aug}

\begin{abstract}
Human mobility describes physical patterns of movement of people within a spatial system. Many of these patterns, including daily commuting, are cyclic and quantifiable. These patterns capture physical phenomena tied to processes studied in epidemiology, and other social, behavioral, and economic sciences.
This paper advances human mobility research by proposing a statistical method for identifying locations that individual move {\em to and through} at a rate proportionally higher than other locations, using commuting data for the country of New Zealand as a case study. These locations are termed {\em mobility loci} and they capture a global property of communities in which people commute. The method makes use of a directed-graph representation where vertices correspond to locations and traffic between locations correspond to edge weights. Following a normalization, the graph can be regarded as a Markov chain whose stationary distribution can be calculated. The proposed permutation procedure is then applied to determine which stationary distributions are larger than what would be expected, given the structure of the directed graph and traffic between locations. The results of this method are evaluated, including a comparison to what is already known about commuting patterns in the area as well as a comparison with similar features.
\end{abstract}


\end{frontmatter}

\section{Introduction and Background}

Patterns of human mobility describe the movement of individuals or the aggregate movement of groups of individuals over time. One class of aggregate movement is {\em commuting} patterns. That is, the movement of individuals from their home locations to work (most often in the morning), as well as their movement from work back home (most often in the evening). For a given area and for regular working days (weekdays, not holidays) these patterns are periodic, they can be quantified, and analyses can be used to understand them.

There is a rich history of research on understanding human mobility patterns from various disciplines including geography, sociology, anthropology, physics, and demography. \citet{barbosa2018} provide an overview of some earlier approaches as well as recent advances. Research in this domain was reinvigorated in the early 21st century with the increased availability of geo-referenced data at the same time when  ``big data'' was emerging as it's own area of study. A popular source of data for human mobility modeling was based on cell phone telemetry data. \cite{gonzalez2008} were among the first to make use of individual-level cellular phone data to show that individuals move in regular spatio-temporal cycles that can be modeled. \cite{song2010} provided both models for human mobility using a L{\'e}vy process as well as a characterization for the model, showing a high degree of predictability in the daily patterns of individuals. This development continued with \cite{alessandretti2020} showing regional, spatial areas in which movement is contained as well as \cite{yuan2016} who explore the distribution of ``human activity spaces" and demographic differences in these distributions.

While development of methods for understanding human mobility is interesting on its own, it is also fundamental to areas of research including social science, epidemiology, and infrastructure planning. Epidemiology in particular has recently made use of mobility methods with \cite{gilani2020} using mobility to both validate existing pollution exposure models as well as identify demographics for which exposure estimates are heavily biased -- a question previously posed in \cite{park2017}. Other examples, like \cite{alessandretti2022, bonaccorsi2020, kane2021, kraemer2020}, highlight the need to incorporate mobility methods into those of existing outbreak models to better evaluate the effect of interventions in the current COVID-19 pandemic.


One of the biggest barriers to mobility research has traditionally been acquiring high-quality data. Most mobility data are derived from cell-phone or cell-phone-application data collected by telecommunications companies or large technology companies. Telecommunication data are often proprietary and difficult to obtain. Technology company location data are often purchasable, but are generally biased toward users of the application or owners of devices making the generality of analyses based on these data to the larger public difficult to assess.

Human mobility has been studied using various quantitative approaches including the use of graphs or networks where vertices correspond to spatial regions and edges capture some aspect of movement between those regions (e.g. \cite{hossmann2011}, \cite{ruan2019}, \cite{chen2022}). These graphs can be referred to as {\em mobility networks} and are distinguished from other types of networks (transportation networks, infrastucture networks, etc.) in that they capture human movement regardless of physical infrastructure facilitating transportation, including roads \citep{barthelemy2011}. Mobility networks are slightly more general and, because the method presented will work for either, we will prefer the more general term.

Methods for analyzing mobility networks focusing on properties of vertices allows us to determine various characteristics of the corresponding spatial locations. Vertex properties can be ``local'' meaning that they can be derived from a vertex and its neighborhood (a subgraph including the vertex). These properties include in-degree, out-degree, whether or not a vertex is a sink, whether or not it is a source, etc. Vertex properties can also be ``global'' meaning the property is not local. These properties include eigenvector centrality, graph radius, graph diameter, whether or not a vertex is a central point, etc. 

In this paper, we propose a new global vertex characteristic of a directed graph identifying ``mobility loci" or simply loci, which are vertices that act as ``hubs" of movement. Vertices designated as loci are those that attract movement \emph{to and through} them in greater proportion compared to other vertices. These loci are identified by accounting for the global properties of the graph in which they reside. This is a statistical property found by estimating a Markov transition matrix from the graph, calculating the stationary distribution of the resulting Markov chain, and then testing to see which vertices have stationary distributions greater than expected under a null distribution sampled from a permutation procedure.


To showcase this procedure, we consider an application to aggregate commuting patterns for the country of New Zealand based on census data. These data have the advantage that they are relatively unbiased with respect to the populations being sampled (commuters). Spatial groups are defined by Statistical Area 2 (SA2), which partitions the country into areas that are comparable in terms of population and other factors. These data are publicly available making them accessible both through the supplemental materials provided in this paper as well as through the Stats NZ Tatauranga Aotearoa, New Zealand's official government data agency. The process for curating these data and building the mobility graphs could be repeated for other countries providing similar census data.

This paper proceeds by providing a more complete description of the New Zealand commuting data, including a description of how routing data were derived, and a brief visual exploration. Subsequent sections interleave statistical and mathematical concepts with their application to these data and culminate in the procedure of calculating mobility loci alongside their identification for the country of New Zealand. The intention for this format is to both construct a statistical procedure as well as provide insights to better understand commuting patterns in the country. Section 2 provides an overview of the data, a description of the preprocessing required to go from raw census data to directed mobility graphs, and an overview of the spatial properties of the resulting mobility graph. Section 3 characterizes the directed mobility graph and constructs the optimization for finding the stationary distribution. We note that the calculation of stationary distributions is not new and can be found in various sources, mostly online or in the waypoint literature \citep{navidi2004,hyytia2006, mitsche2014}. However, for completeness, we have roughly followed \cite{chang2013} before providing a formal construction. Section 4 proposes a test for finding loci, those elements of the stationary distribution that are larger than what is expected when keeping the graph structure fixed and permuting on traffic between SA2 areas, as well as a procedure for finding groups of loci while addressing multiple-testing challenges. Section 5 casts the stationary distributions and loci as global vertex properties. Because the notion of mobility loci is a novel vertex property, there is not a direct comparison with other procedures to evaluate the results. However, we do provide analyses quantifying how much information is encoded in these features compared to local graph features including vertex degree and others explained later. Section 6 includes potential applications for this work, which fit readily into the spatial research framework as well as other potential application areas.

\section{New Zealand Commuting Data}

\subsection{Data Overview}

The New Zealand 2018 Census includes among other questions information about the main means of travel to work. Based on the answers to this question and respective residence/workplace addresses the Stats NZ Tatauranga Aotearoa, New Zealand's official data agency, publishes a commuter view dataset \citep{nz.data} which aggregates the number of usually resident population aged 15 years and over by main means of travel to work. The spatial aggregation is done by ``statistical areas"  \citep{nz.sa} which is a spatial partition of the entire country with focus on retaining comparably constant population per partition in urban and suburban areas. The commuter dataset uses partitions induced by Statistical Area 2 (SA2) polygons (represented as shape files) which typically have population of 2,000 to 4,000 residents in urban areas.

The data use fixed random rounding to protect confidentiality. Counts of less than 6 are suppressed according to 2018 confidentiality rules \citep{nz.conf.rules}. For the purpose of this analysis we will ignore suppressed values which may lead to a slight under-count in sparsely populated rural areas, but does not affect urban or sub-urban areas. Given our additive treatment of the individual counts we expect the rounding to not have a significant impact given the magnitude of the resulting values.

\begin{figure*}
\centering
\begin{tabular}{cc}
\includegraphics[width=0.7\textwidth]{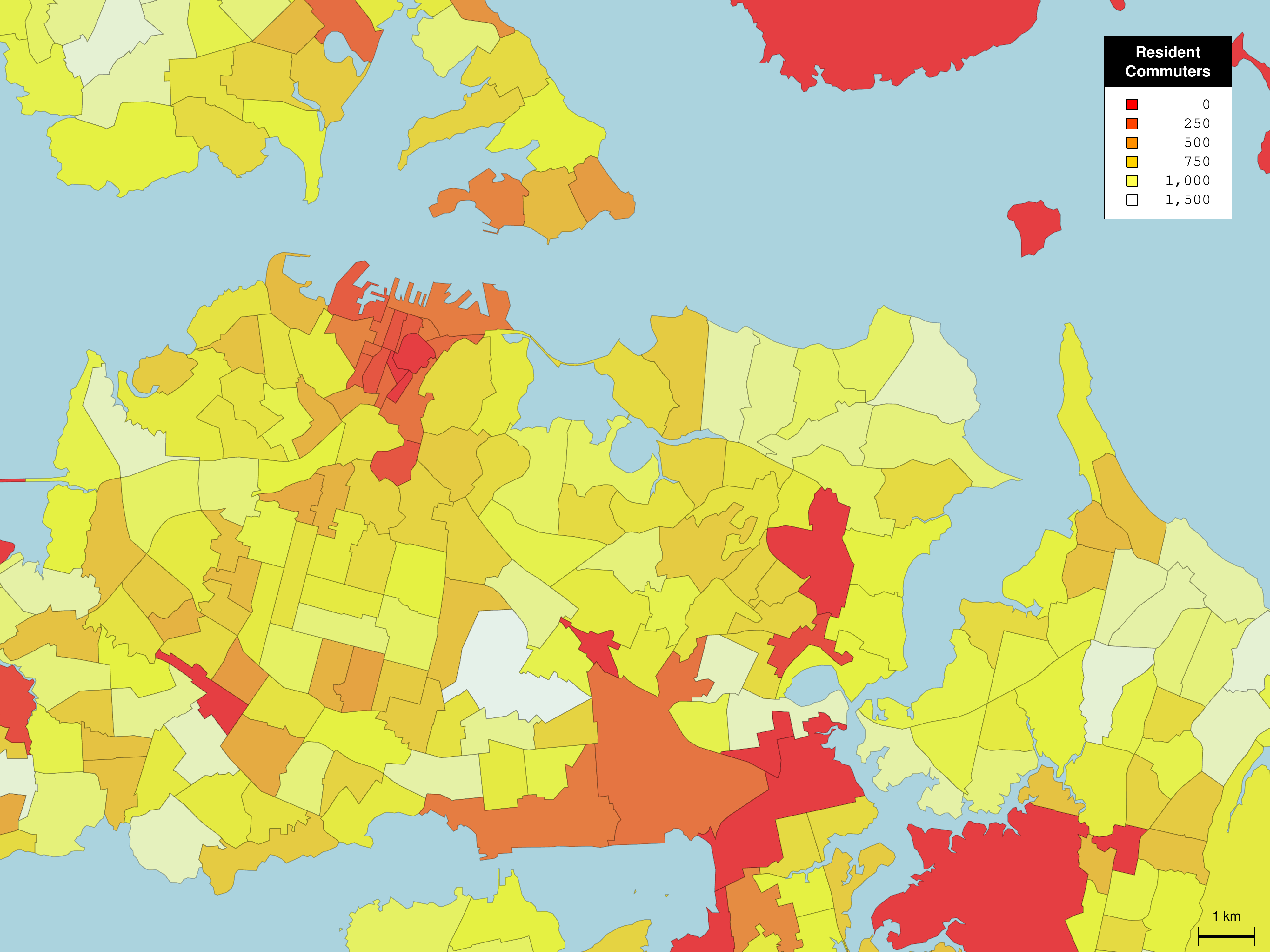} & (a)\\
\includegraphics[width=0.7\textwidth]{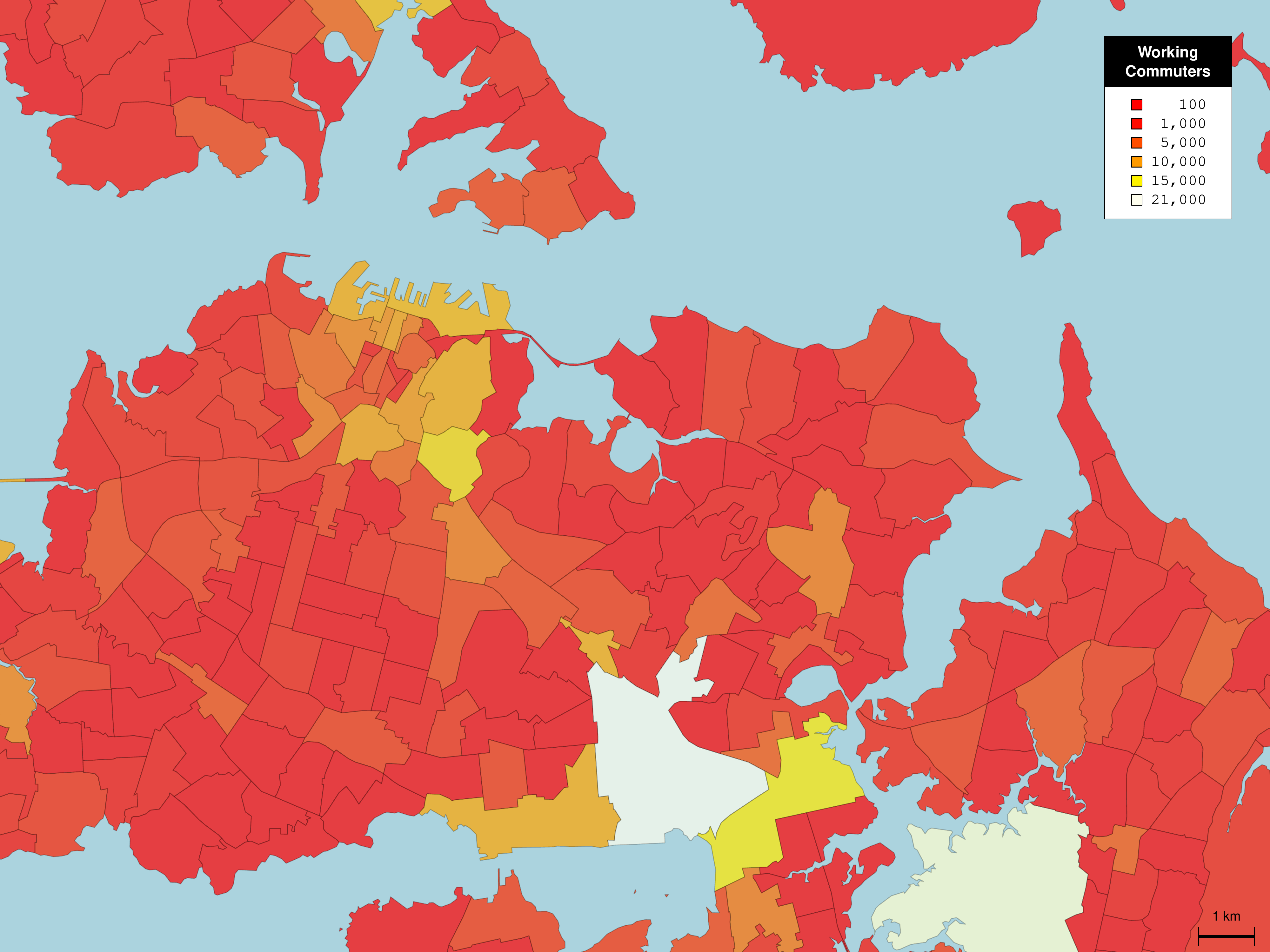} & (b)
\end{tabular}
\caption{ 
\textbf{(a)} Spatial distribution of commuter residential locations
\textbf{(b)} Spatial distribution of commuter work locations
}
\label{fig:spatial-counts}
\end{figure*}

The census data, as provided, include SA2 of usual residence as well as those of the workplace. They do not include the actual routes taken by individuals on particular days. Figure~\ref{fig:spatial-counts} shows (a) the spatial distribution of commuter residential locations while (b) shows the spatial distribution of commuter work locations, with white indicating the highest counts and red denoting the lowest. These two maps show that residences are evenly dispersed across the greater Auckland area while work locations are relatively concentrated. They also reveal that the spatial distributions for residential and work locations are fairly inverse of one another - high density residential areas have lower work location densities. The red-orange colored cluster of SA2s on the water in the center-west of figure \ref{fig:spatial-counts} (a) (with small areas) constitutes Auckland's Central Business District, New Zealand's leading financial hub and the centre of the country's economy. The white SA2 in Figure~\ref{fig:spatial-counts} (b) is part of the Penrose district, an industrial suburb.  Unsurprisingly, there are relatively few residents in these commercial areas. 

\subsection{Data preprocessing: Constructing directed mobility graphs}

The goal of the preprocessing step is twofold. The first is to determine the most likely route taken by commuters from their residential SA2 area to their work location. We note that on a given day the route taken to work may vary based on many factors including, construction, congestion, etc. and that many residents may not take the most direct or shortest route for any number of reasons such as carpooling, dropping children off at school, etc. However, we assume that the procedure used is sufficient to capture commuting paths for most of the residents enough of the time for the analysis to be valid. 

In order to model actual movement through space, we focus on the commutes using personal vehicles. The \citep{ghopper} routing engine was used 
to infer the most likely route from home to work, and road topology was sourced from Open Street Map \citep{osm}. The resulting route is represented as a line segment, which is then used to infer the sequence of SA2s along the way. One additional complication is the fact that SA2s often use roads as boundaries so small deviations may cause apparent frequent movement between adjacent SA2s along the road. To counteract that effect we use 10 meter buffers around borders and will consider a transition from one SA2 to another only if the route has fully left the SA2 including the extra margin. Sequences can be reversed for evening commutes and the same approach can be used for different modes of transport (public transit, cycling, etc.).




The second goal of the preprocessing step is to aggregate the individual-level commuting sequences into a directed graph. SA2 areas are represented as vertices in the graph. Any movement from one SA2 to another constitutes a directed edge. The weight of the edge is the count of transitions between pairs of SA2s. Note that by design, edges can only lead from one SA2 to its neighboring SA2s.

\begin{figure*}
\begin{center}
\includegraphics[width=0.9 \textwidth]{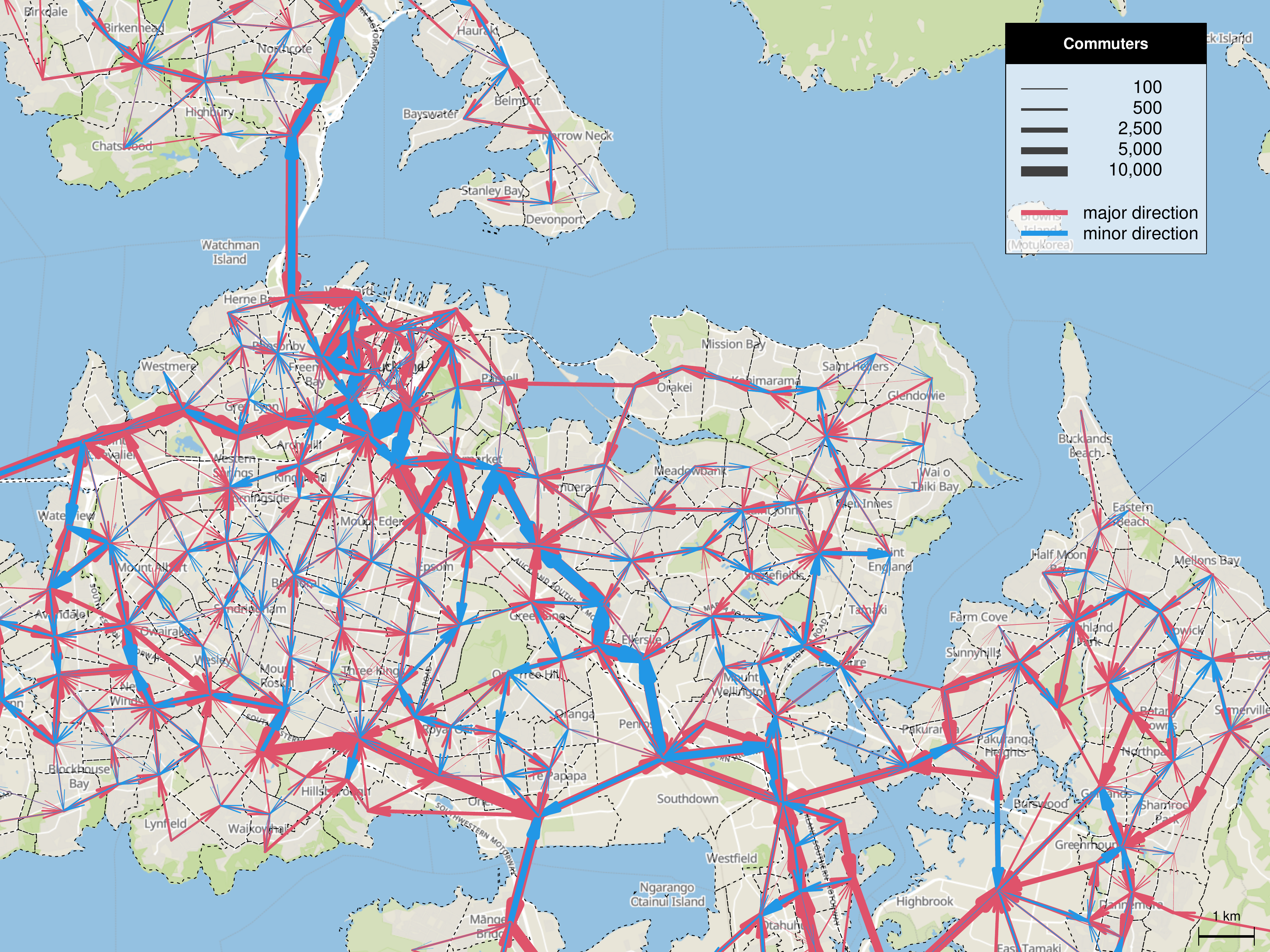}
\end{center}
\caption{Commuter movement between SA2 areas at commute time from home to work in the central Auckland region. Edges are drawn from the area centroid. The thickness of the directed edges is proportional to the square root of the number of people moving from one area to another, while the arrows indicate the direction. The major direction of movement has been drawn in red and minor in blue, therefore edges with significant blue thickness have relatively high counter-movement against the major direction. \label{fig:plot2}}
\end{figure*}

A small subset of the resulting graph corresponding to the region around central Auckland is illustrated in Figure~\ref{fig:plot2}. The graph represents the commute from home to work, typically corresponding to the morning commute. Each arrow represents the directed movement from one SA2 to another. In many cases there is a movement in both directions for a pair of areas, where some people leave one area and others enter that area. In order to visually distinguish the magnitudes of the flows, we define a ``major'' direction which for a pair of vertices is the edge with the higher weight. In Figure~\ref{fig:plot2} we color the major direction edges in red and correspondingly the opposite minor direction (if it exists) in blue.

The total width of the red arrows shows us the general major movement between regions. Movements that are counter to the major direction are then seen in blue. We can see an entire series of such counter-movement in the center of the plot, corresponding to the main artery in the region: State Highway 1. The presence of those large counter-movements should be noted because it contradicts a common assumption that workers move from suburbs into the city for work in a ``spoke-and-hub'' fashion. This is not supported by the visualization. There is a significant north-south flow in both directions, while east-west flows dominate in the direction towards the city.

\begin{figure*}
    \centering
    \includegraphics[width=.9\textwidth]{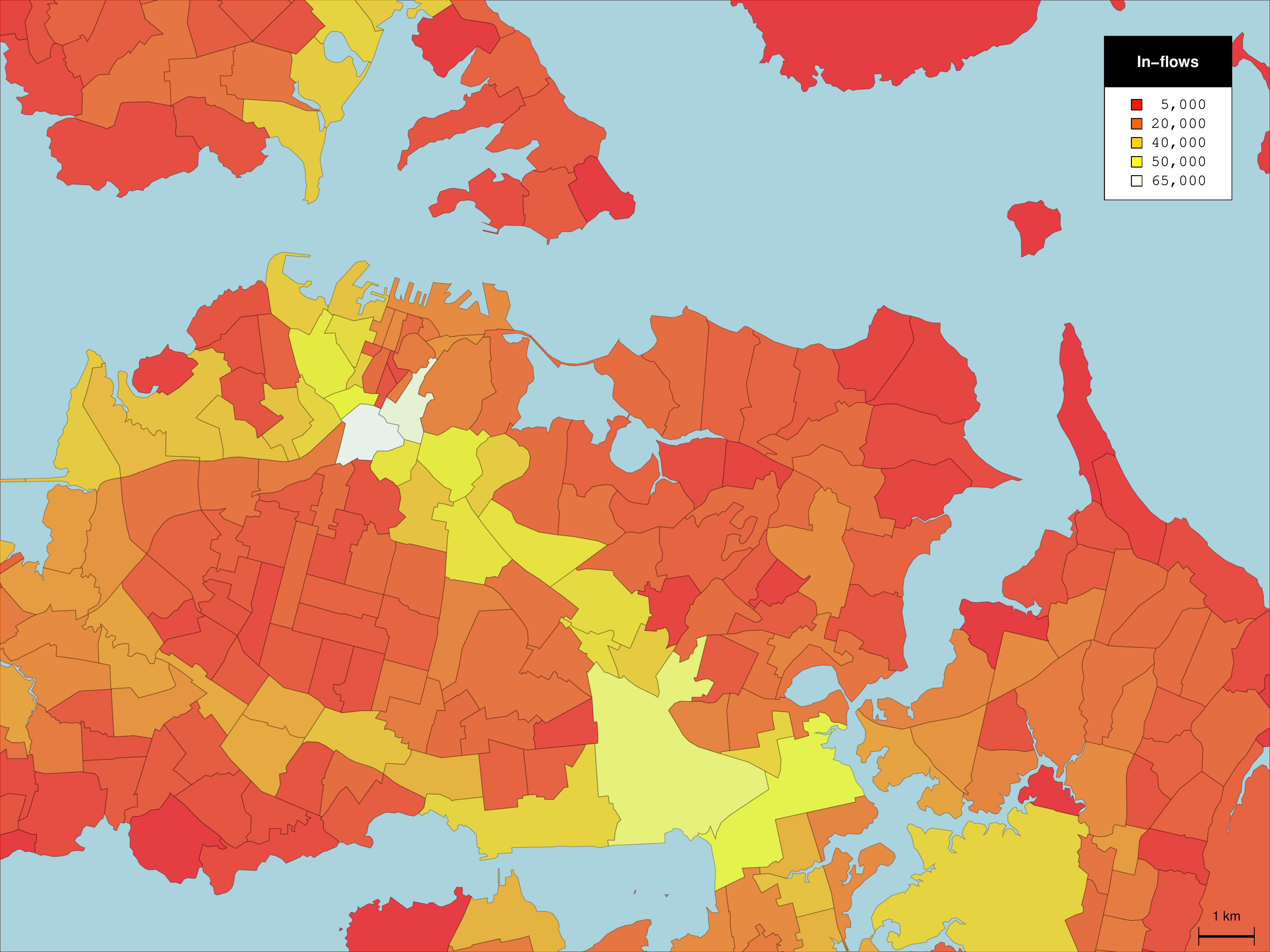}
    \caption{Distribution of the traffic in-flows, the count of all individuals who enter an SA2, regardless of whether they stay there or subsequently leave.}
    \label{fig:map-flowdist}
\end{figure*}


Figure~\ref{fig:map-flowdist} shows the distribution of the in-flows with white indicating the highest counts and red denoting the lowest. In-flows count all individuals who enter an SA2, regardless of whether they stay there or subsequently leave. The figure reveals that, as expected, contiguous areas with high traffic are associated with major highway corridors.




\section{The directed mobility graph}

\begin{figure*}[htbp]
\begin{center}
\includegraphics[width=0.7 \textwidth]{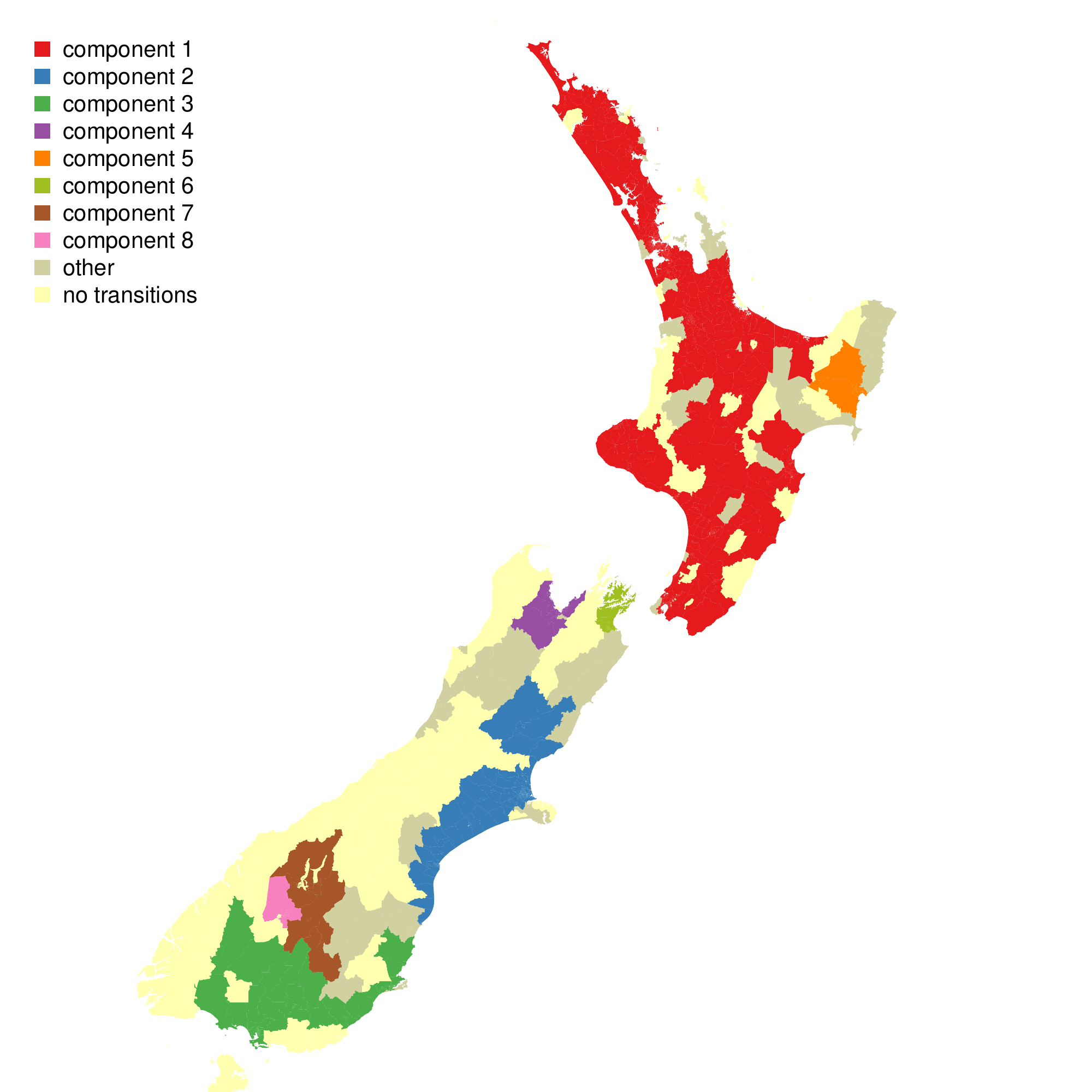}
\end{center}
\caption{Strong components in the New Zealand commuter mobility graph for the morning commute.}
\label{fig:plot3}
\end{figure*}

The result of the preprocessing step on the morning commute data is the construction of a {\em directed mobility graph} that encodes the SA2s as vertices and the aggregate movement between adjacent spatial areas as edge weights in a directed graph. This can be thought of as an extension to the origin-destination matrix described in \cite{willumsen2002}, containing not only start and endpoints but the intermediate transitions as well. We then  partition the directed graph into strong components, i.e., sets of vertices that can be reached from any other vertex in the component. Figure~\ref{fig:plot3} shows these strong components. The largest one, in red (component 1), contains 1426 SA2 areas covering most of the North Island and has the largest population, and will be the focus for the rest of this analysis.

\begin{figure*}
\begin{center}
\includegraphics[width=0.7 \textwidth]{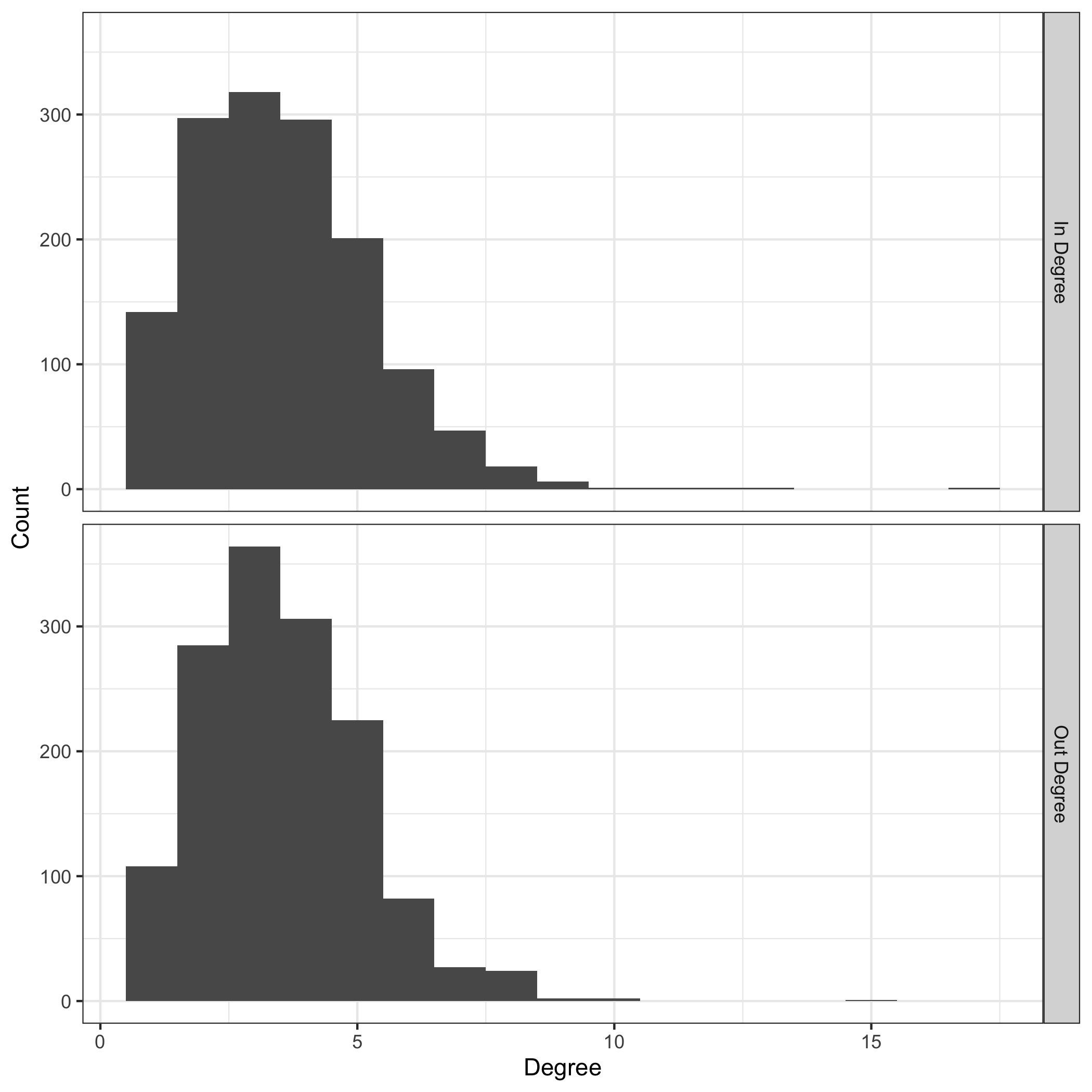}
\end{center}
\caption{The histogram of the edge weights in the largest strong component of the mobility graph.}
\label{fig:hists}
\end{figure*}

Histograms of the edge weights in the largest component appear in Figure \ref{fig:hists}. The weights appear to decrease exponentially, indicating few transitions between most adjacent SA2 areas, except for a few. This is likely because a large portion of the SA2 areas correspond to rural areas, where there are fewer transitions in general as well as those SA2 transitions in more suburban and urban areas with populations who are not commuting to high population-density areas, like city centers.

\begin{figure*}[htbp]
\centering
\begin{tabular}{cc}
\includegraphics[width=0.45\textwidth]{figures/degree-hists.png} &
\includegraphics[width=0.45\textwidth]{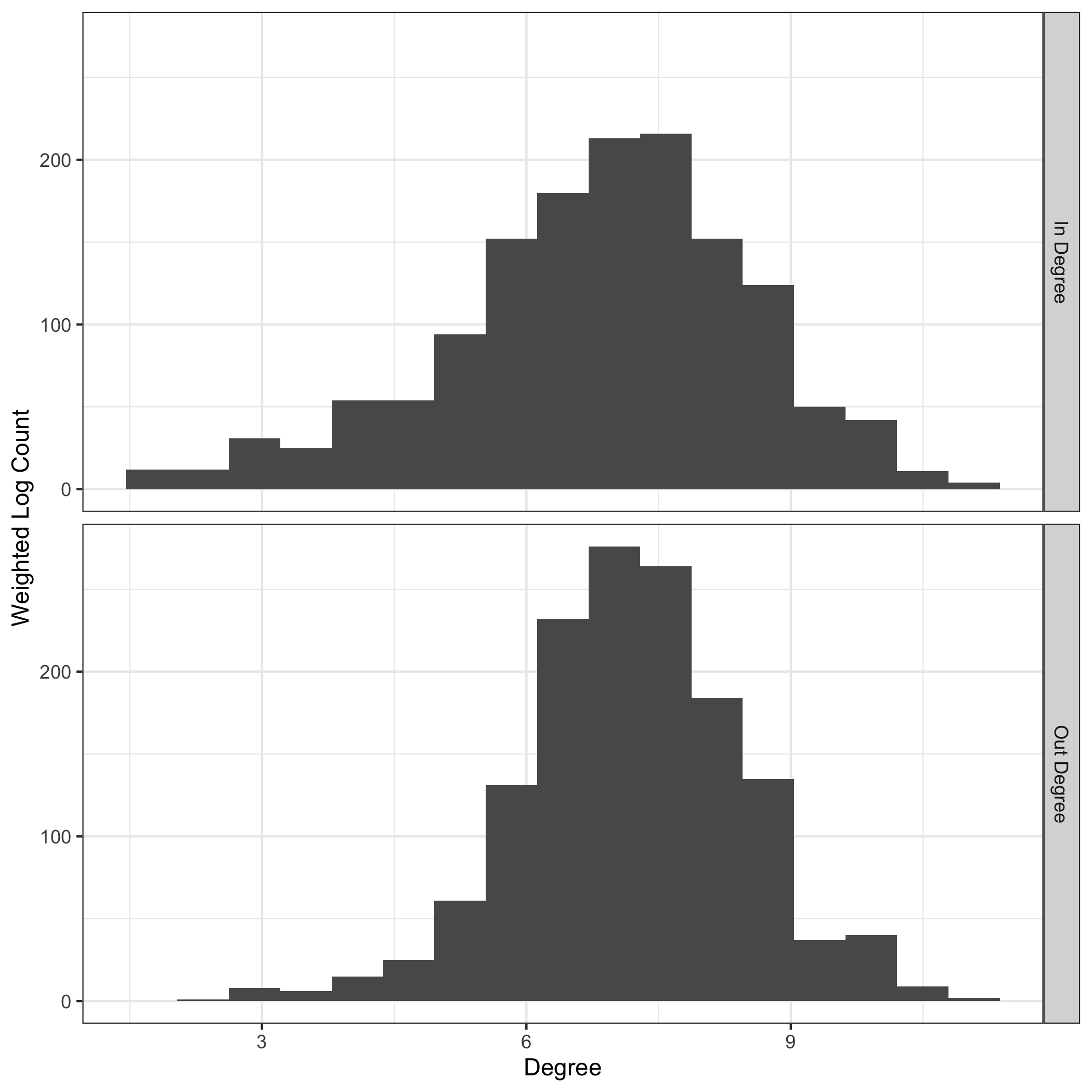} \\
\textbf{(a)}  & \textbf{(b)}  \\[6pt]
\end{tabular}
\caption{ 
\textbf{(a)} The histogram of in- and out-degrees for morning commutes of the largest strong component.
\textbf{(b)} The histogram of log of edge-weighted in- and out-degrees for morning commutes of the largest strong component.
}
\label{fig:degree-dist}
\end{figure*}

Figure \ref{fig:degree-dist} (a) shows histograms of the in- and out-degrees vertices in the largest component. These two distributions are relatively symmetric with the out-degree histogram having a slightly larger mode and the in-degree histogram having a slightly heavier tail. Since these are taken from a strong component and each vertex must have at least one in-edge and one out-edge we may conclude that, on average, vertices incoming and outgoing traffic is balanced.
Figure \ref{fig:degree-dist} (b) shows a histogram of the log of the edge-weighted in- and out-degrees. The out-degree distribution is sightly more concentrated than that of the in-degree. This is likely an artifact of the difference between rural/suburban and urban SA2 areas, where in rural/suburban areas, a greater proportion of commuters travel in the direction of more urban SA2 areas and have a larger weighted out-degree. Urban areas tend to have more symmetric weighted in- and out-degree values.

\subsection{Directed mobility graph to stationary distributions}

The directed mobility graph with $n$ vertices can be represented as a matrix, $M \in \mathbb{R}^{n \times n}$, quantifying the aggregate movement between adjacent areas with the rows corresponding to the ``from'' (origin) location, columns corresponding to the ``to'' (destination) locations, and elements of the matrix corresponding to the number of people moving between respective locations. The sum of a column captures the total movement of people to a location.

Let $P \in \mathbb{R}^{n \times n}$ be the matrix that results from normalizing over the rows of the mobility graph and let $P_{i, \cdot}$ denote the $i$th row of $P$.
$$
P_{i, \cdot} =  M_{i, \cdot} / \sum_{j=1}^n M_{i,j} 
$$
The scaling provides a mobility measure relative to individuals, rather than total movement between SA2s, allowing us to directly compare movement patterns between areas with different population densities. It is also equivalent to estimating the transition probabilities between SA2 areas using the maximum likelihood estimator, under the first-order Markov assumptions. This matrix will be referred to as the Probability Transition Matrix (PTM). 

\begin{figure*}
\begin{center}
\includegraphics[width=.8 \textwidth]{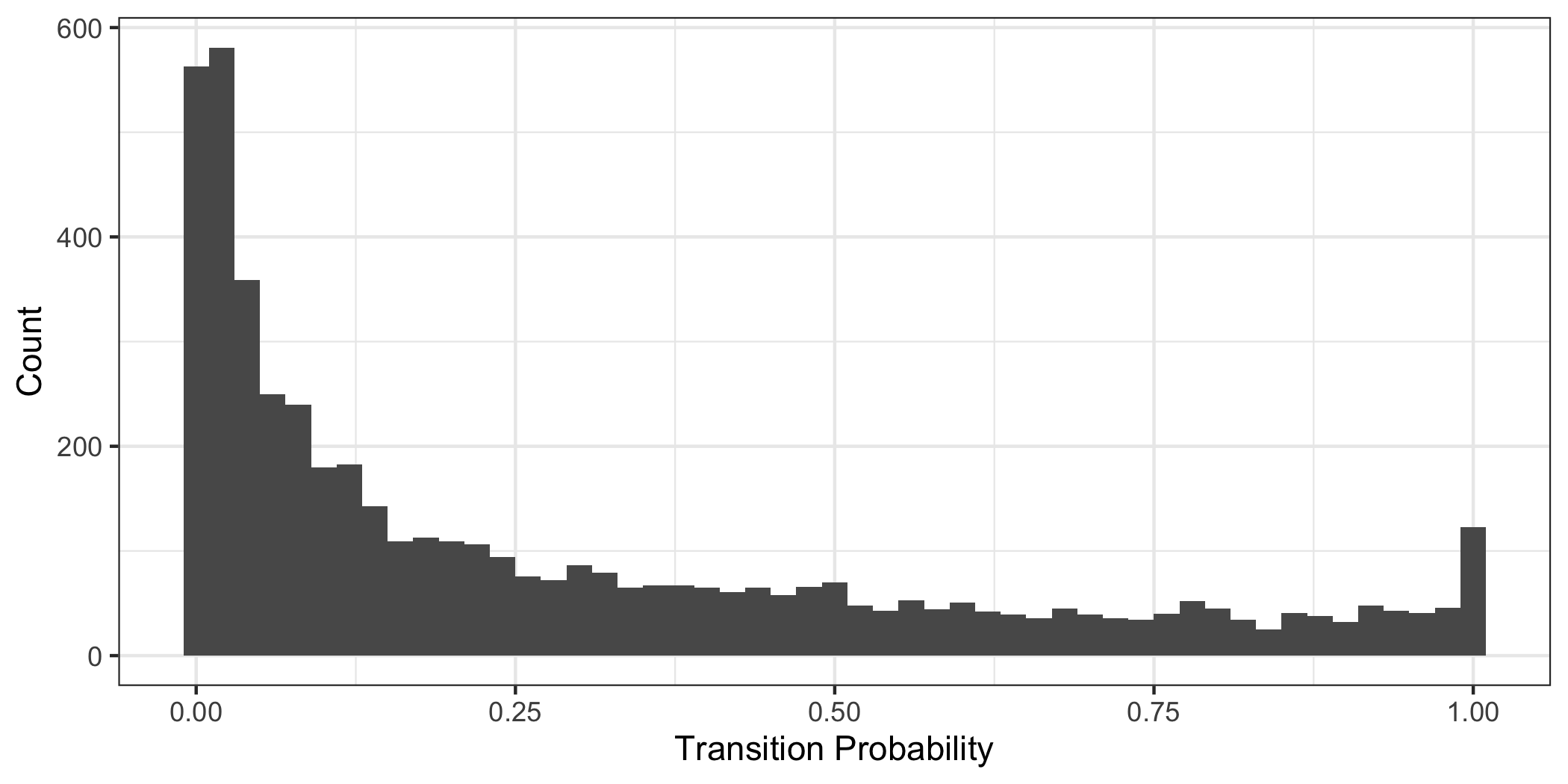}
\end{center}
\caption{The histogram of values of the non-zero transition probabilities.}
\label{ptm-hist}
\end{figure*}

Supplemental \ref{ptm-hist} shows the histogram of the non-zero transition probabilities for the largest strong component in the New Zealand data. There are several things to note. First, this visualization is for both morning and evening commutes since the morning commute PTM is the transpose of that of the evening commute. Second, the counts drop off roughly exponentially. This may be because a large number of people live in suburban areas, outside of the city. The first few transitions for these commuters can vary drastically, depending on their destination, corresponding to  low-probability transitions. As they transition to secondary or primary roads, ``traffic flows'' become more regular with less variation on commuting patterns, corresponding to higher transition probabilities on fewer roads. Third, there is a spike at transitions near a value of one. This likely corresponds to traffic along highways outside of urban areas where individuals are commuting toward urban areas.

\subsection{Calculating the stationary distribution} \label{calc_stat_dist}

The PTM is an object allowing us to create a procedure for evaluating the aggregate movement of individuals in between SA2s as a Markov Chain. To quantify mobility with these data, we propose a feature based on the stationary distributions of SA2 areas, which is defined as the probability distribution
\begin{equation} \label{eqn:stat_dist}
    \pi = P \pi,
\end{equation}
where $P$ is the PTM. In order to be able to calculate the stationary distribution, and for it to be unique, two conditions on $P$ must be satisfied \citep{chang2013}. First, the matrix $P$ must be {\em irreducible}, meaning that there is a path from any area to any other area in the mobility graph. Second, $P$ must be {\em aperiodic}, meaning that the greatest common divisor (gcd) of $\{m : (P^m)_{i,i} > 0\}$ for each $i \in 1, ..., n$.

The first is satisfied by conditioning on strong components in the mobility graph, which ensures that there is a path from each vertex to any other vertex in the strong component. The second can be checked directly by examining the diagonals of $P^k$ for {$1 \leq k \leq n$} since the strong component condition guarantees that the path length from a vertex to itself has length less than or equal to the number of vertices in the strong component. The gcd can then be calculated, for each of the diagonal elements of $P^k$. If the gcd of any of those values is not 1, then the PTM is periodic and convergence is not guaranteed.


When these two conditions are met, we can solve for $\pi$ directly by first rearranging the terms
\begin{align*}
    0 & = P \pi - \pi \\
    0  & = P \pi - I \pi \\
    0  & = (P - I) \pi
\end{align*}
for the identity matrix $I \in \mathbb{R}^{n \times n}$. We now have a linear system of $n$ equations of $n$ variables. To constrain $\pi$ so that its sum is one, we add the following row to the system.
\begin{equation*}
    \begin{bmatrix}
    0 \\
    1
    \end{bmatrix} = 
    \begin{bmatrix}
    P - I\\
    \pi
    \end{bmatrix}.
\end{equation*}
This system of equations, plus the constraints that the value of $\pi$ must be at least zero and at most one, define a constrained, linear-optimization problem whose solution can be found via standard methods. 

There are two interpretations for the stationary distribution in this setting. The first, standard statistical interpretation is that if an individual were to start at a random SA2 and proceed to an adjacent SA2 according to the probabilities in the PTM, then the stationary distribution is the proportion of time the individual visits each SA2 area as time goes to infinity. Another related interpretation is that the chain is a random dynamical system, with discrete time and discrete state space. The stationary distribution $\pi$ is a {\em fixed-point attractor} since it is mapped to itself by the PTM as described in Equation \ref{eqn:stat_dist}. Roughly, we can interpret SA2s with relatively large stationary distributions as areas individuals tend to \emph{go to and through}, regardless of their starting point.

\section{A permutation test for loci}

The attractor, as defined in this paper, is $\pi$, the vector of stationary distributions with index corresponding to SA2 area. The attractor's values depend on two properties of the mobility graph. First is the graph structure, which is characterized by the connectivity between vertices, encoded with the edges. Second is the individual-level preference for direction, which is encoded by the edge weights as transition probabilities. The stationary distribution of SA2's show that some elements have values much larger than others. However, these large observed values might be unremarkable given the structure of the graph. Based on this observation a natural question to ask is, 
``which attractor elements have large values that are greater than expected, {\em given the structure of the mobility graph for New Zealand's largest North Island strong component}?'' Posing the question in this way allows us to identify those SA2 areas that individuals tend to go to and through, independent of the overall population of the area or the total number of people going to or through an area, at a rate that is unusually higher than expected. We term such an SA2 area a \emph{locus}.

Samples from the null distribution of attractors can be derived by randomly permuting the edge weights of the mobility graph, and then calculating the stationary distribution. Fixing the structure of the graph tailors the distribution to that of the mobility graph, while permuting over edge weights fixes the {\em total movement} captured by the mobility graph. We will refer to the null attractor distribution as $\Pi$. Let $\Pi_i$ and $\pi_i$ denote the  $i^{th}$ element of $\Pi$ and $\pi$ respectively, and let $\alpha$ be a suitable cutoff. Then, we formally define the element $\pi_i$ as a {\em locus} if 
\begin{equation} \label{eqn:pprob}
    \mathbb{P} \{ \Pi_i \geq \pi_i \} \leq \alpha.
\end{equation}

\begin{figure*}[htbp]
\begin{center}
\includegraphics[width=0.8 \textwidth]{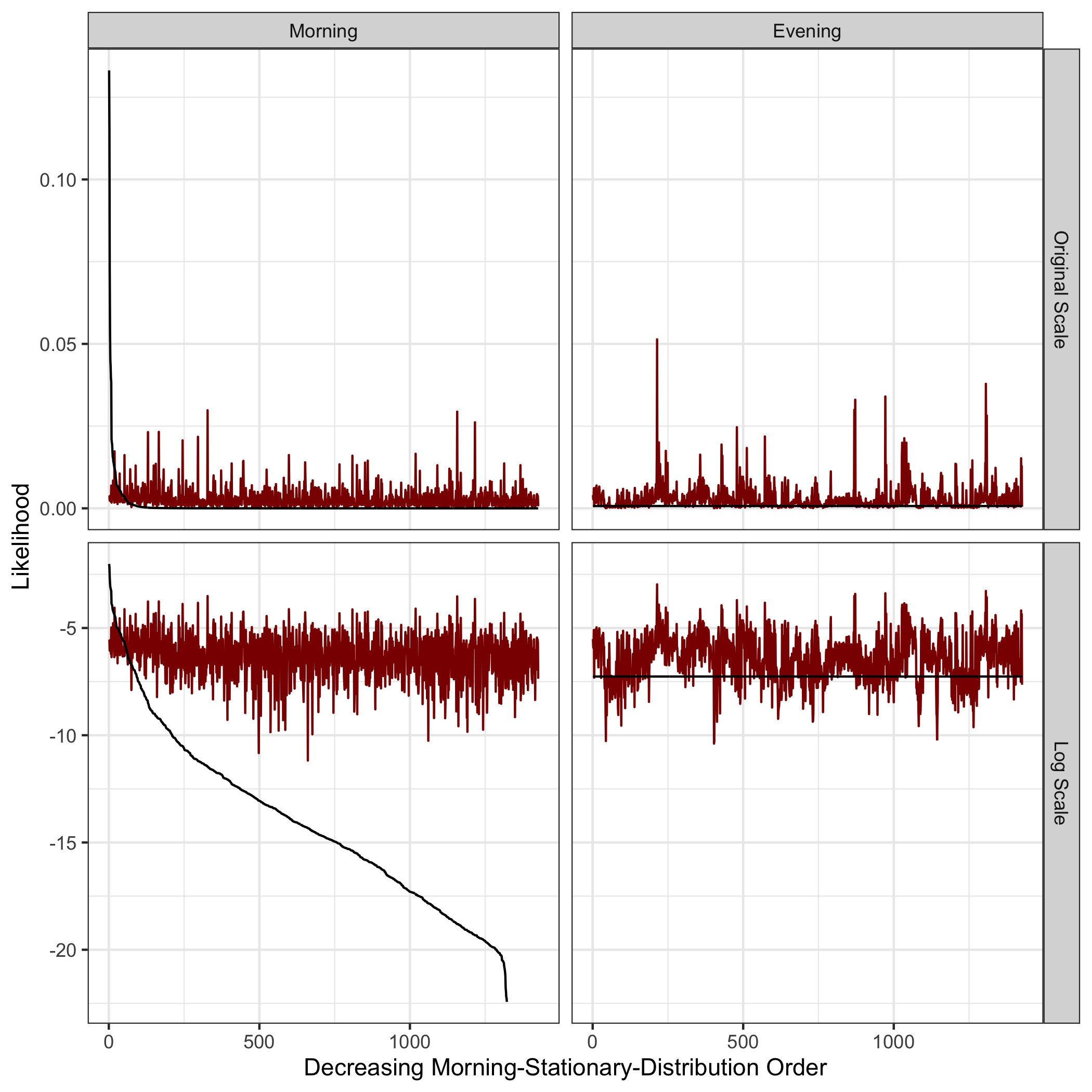}
\end{center}
\caption{The stationary distribution of all vertices, in black, in descending order of the morning stationary distribution, of the morning and evening commutes on the original and log scale along with the quantile values of $\Pi$, in red, when $\alpha = 5$\%.
}
\label{fig:tail-probs}
\end{figure*}

Figure~\ref{fig:tail-probs} shows the stationary distributions, in black, of the morning and evening commmutes on the original and log scale along with the quantile values of $\Pi$, in red, when $\alpha = 5$\%. The values are in descending order of the morning commute stationary distribution. The largest morning stationary distributions start with a value of 0.133 and drops off quickly to values close to zero. The corresponding quantile values are relatively small meaning that it is unlikely that larger stationary distribution values would occur under $\Pi$. The graphs of the evening commutes are quite different. The stationary distributions are all very close to 0.000701, which is the probability mass function of the uniform distribution  counting numbers from one to 1426 (the number of vertices in the strong component). A total of 87 of the stationary distribution values fall below the 5\% threshold. 

\begin{figure*}[htbp]
\begin{center}
\includegraphics[width=.8 \textwidth]{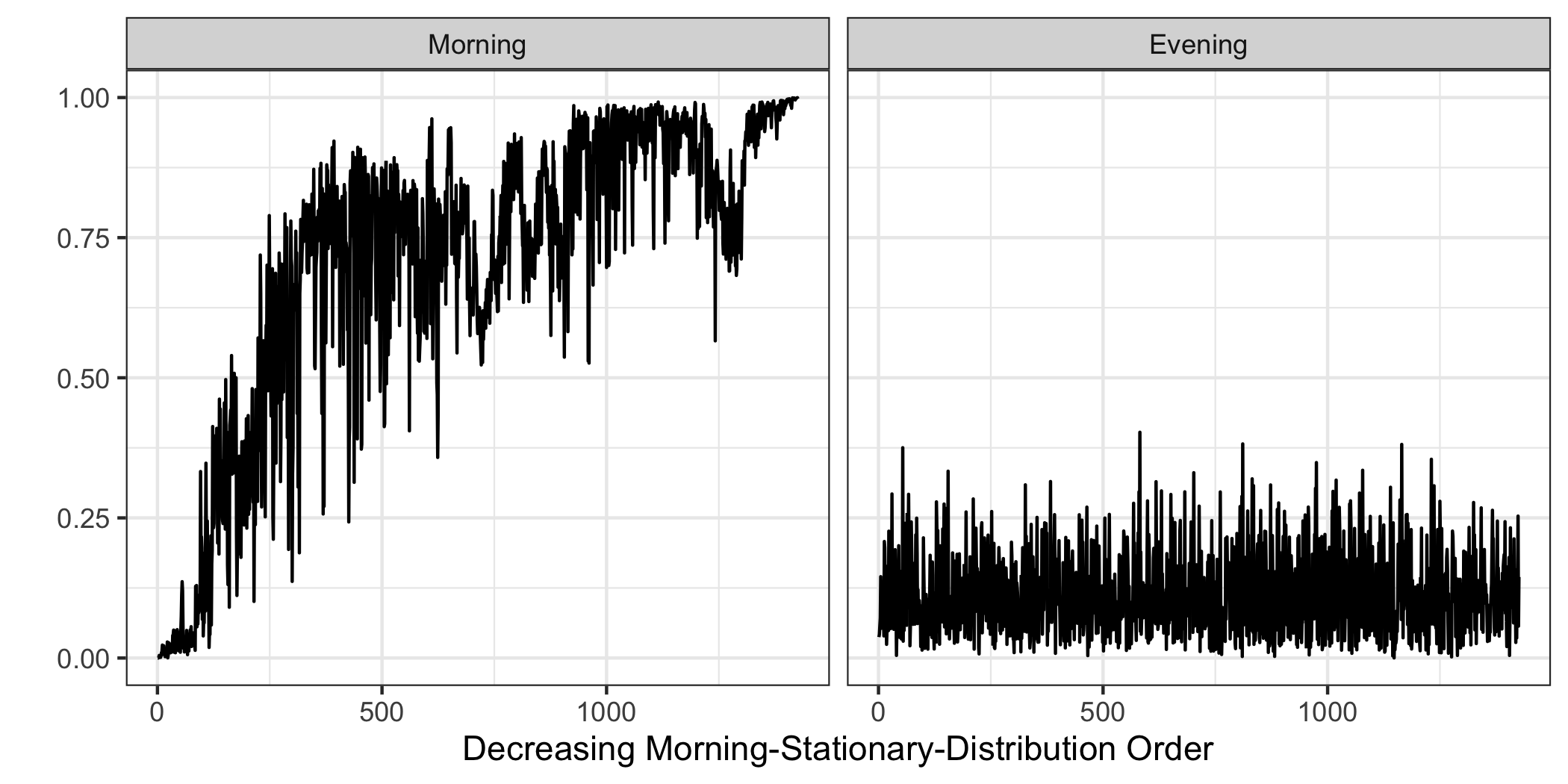}
\end{center}
\caption{The tail probabilities and stationary distribution of all SA2's, ordered by largest to smallest morning stationary distribution.}
\label{fig:p-plot}
\end{figure*}

Figure~\ref{fig:p-plot} shows the individual estimates of the tail-probabilities ($p$-values) for morning and evening commutes. Together, the plots suggest that morning commutes see the concentration of individuals from their home locations to areas with higher than expected stationary distributions ({\em loci}), while the evening commute does not experience this same phenomenon and behaves more closely to uniform mixing over the strong component.

\subsection{A procedure for identifying high-probability SA2 areas for morning commutes}

The previous section essentially amounts to a statistical test where elements of $\pi$ are used to calculate the tail probability (p-value) under the null distribution, which was sampled by a permutation procedure. While this procedure is effective for testing individual elements of $\pi$, identifying sets of loci brings with it the multiple testing challenge. This challenge is mitigated by the fact that, as stated previously, we are interested in the intersection of SA2 areas whose stationary distribution is large {\em and} whose value is greater than expected. For the rest of this section we will focus on morning commutes.

Let $\Pi_{(i)}$ and $\pi_{(i)}$ be the $i^{th}$ values of $\Pi$ and $\pi$ ordered in decreasing value of $\pi$. Let 
\begin{equation}
    X_{(i)} = \mathbb{P}\{\Pi_{(i)} \geq \pi_{(i)}\}
\end{equation}
be the tail-probability of $\pi_{(i)}$ with respect to $\Pi_{(i)}$. Let $k$ be an integer from 1 to $n$, the number of vertices in the strong component, and let
\begin{equation}
    X_{(1:k)} = X_{(1)}, X_{(2)}, ..., X_{(k)}
\end{equation}
be the vector of tail probabilities corresponding to the $k$ largest stationary distributions. This can be thought of as a vector of p-values that can be adjusted for using any of the standard methods for multiple comparisons (\cite{bonferroni1935}, \cite{benjamini1995},  \cite{benjamini2001}, \cite{hommel1988}, etc.) In this analysis we will show results based on  Benjamini-Hochberg procedure. This is the least conservative of the listed methods and the liberalness of the procedure is likely warranted because, as will be shown later, the tests are not independent, with sets of loci tending to be highly spatially correlated.

To find the set of loci for the mobility graph, we find 
\begin{equation} \label{eqn:max-k}
    \argmax_{k} \sum_{i = 1}^k \{ X_{(i)} < \alpha \},
\end{equation}
the $k$ that maximizes the number of loci. We then report those SA2 areas with adjusted tail probabilities less than $\alpha$. We do note that by adjusting over different vector sizes, we are essentially doing a ``test-of-tests.'' However, we remind the reader that the goal of this procedure is not to get the adjusted $p$-values,  but rather it is to maximize the number of loci while taking into account the multiple testing problem for each value of $k$.

\begin{figure*}[htbp]
\begin{center}
\includegraphics[width=.8 \textwidth]{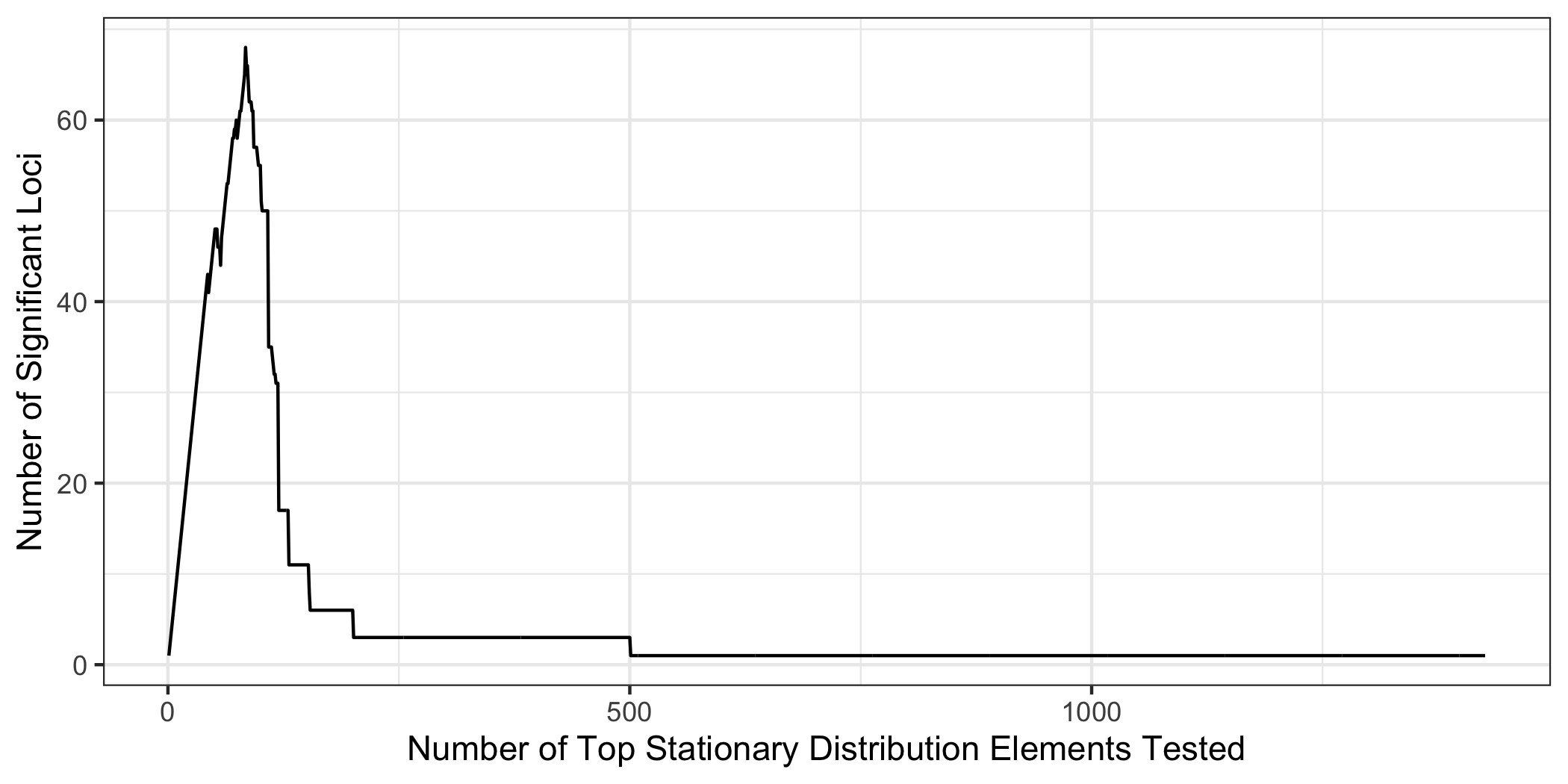}
\end{center}
\caption{The number of significant loci conditioned on top stationary distributions.}
\label{fig:sig-loci-k}
\end{figure*}

Figure~\ref{fig:sig-loci-k} shows the number of significant loci, up to $k$, after adjusting the tail probability ($\alpha = 0.05$) according to the Benjamini-Hochberg procedure. Roughly, as the number of the largest stationary distribution elements increases, so does the number of significant loci. This continues until new stationary distribution values are not enough to justify the increased number of tests and the number of significant loci begins to decrease. However, the increase is not strict and the graph has three local maxima.

\begin{figure*}[htbp]
\begin{center}
\includegraphics[width=.8 \textwidth]{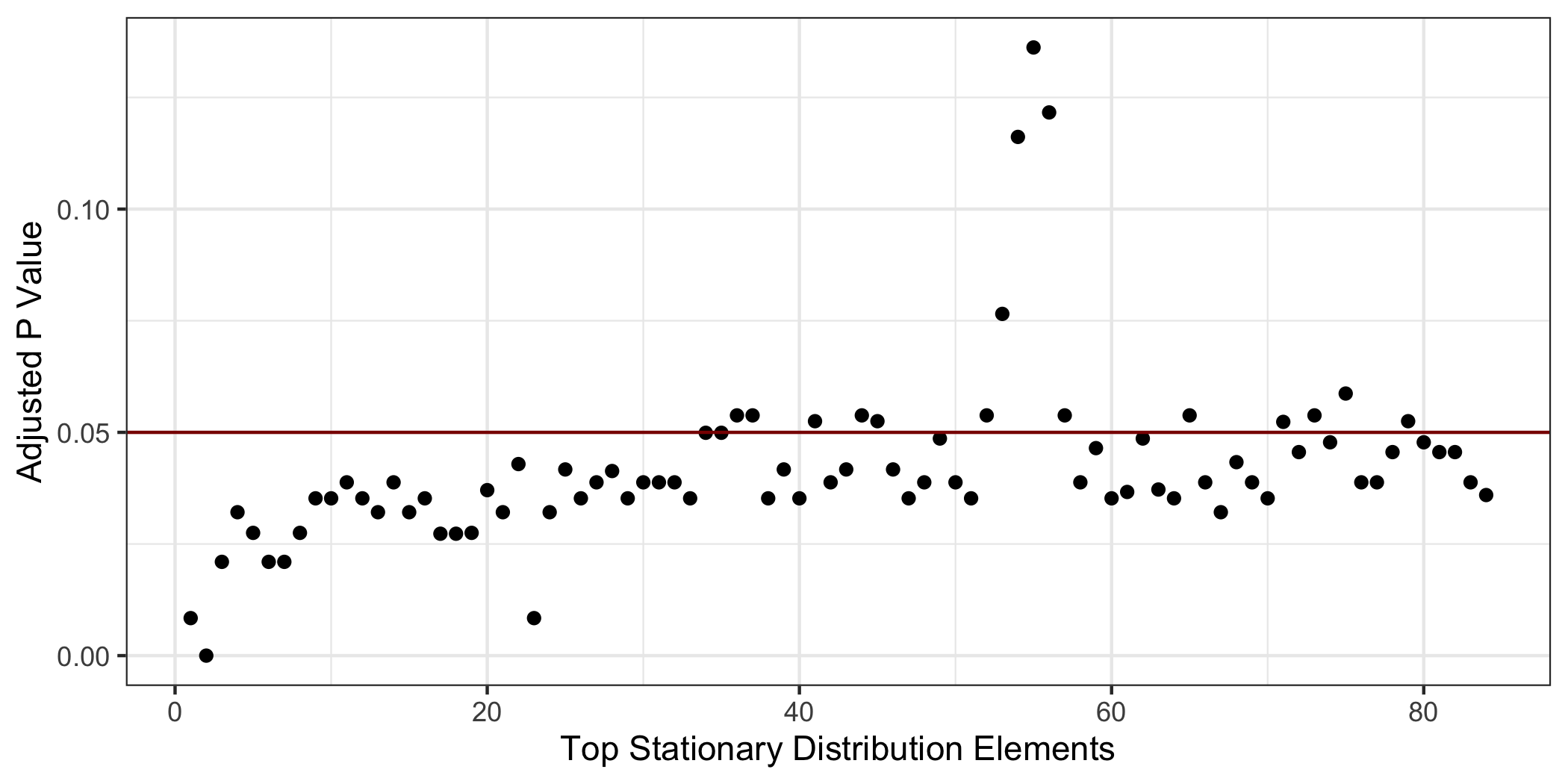}
\end{center}
\caption{The False-Detection-Rate (FDR) adjusted $p$-values.}
\label{fig:adj-p}
\end{figure*}

The value of $k$ maximizing Equation \ref{eqn:max-k} for the largest strong component in the New Zealand mobility graph (with $\alpha = 0.05$) is 68. The graph of the adjusted $p$-values is shown in Figure~\ref{fig:adj-p}. From the graph we can see that it is not true that all of the top stationary distribution values are statistically significant. Depending on the underlying graph structure, we may be able to add more of the top stationary distribution vertices, even when some are not significant, to find the maximum number of adjusted, significant vertices.

\begin{figure*}
\begin{center}
\includegraphics[width=0.9 \textwidth]{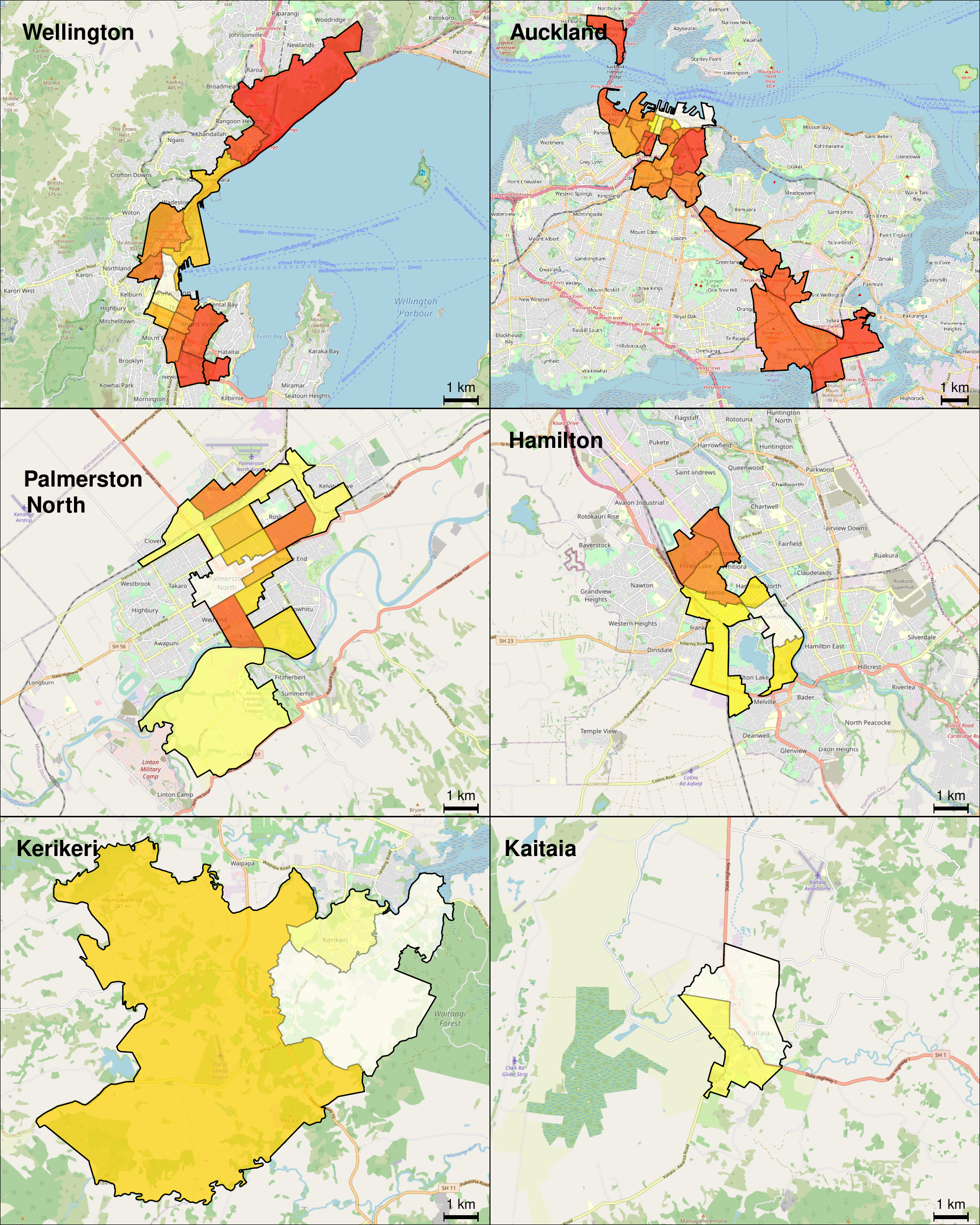}
\end{center}
\caption{Loci in the various regions of the North Island, colored by stationary distribution values (yellow corresponds to larger stationary distribution values).
} 
\label{fig:loci-maps}
\end{figure*}

\begin{figure*}
\begin{center}
\includegraphics[width=0.8 \textwidth]{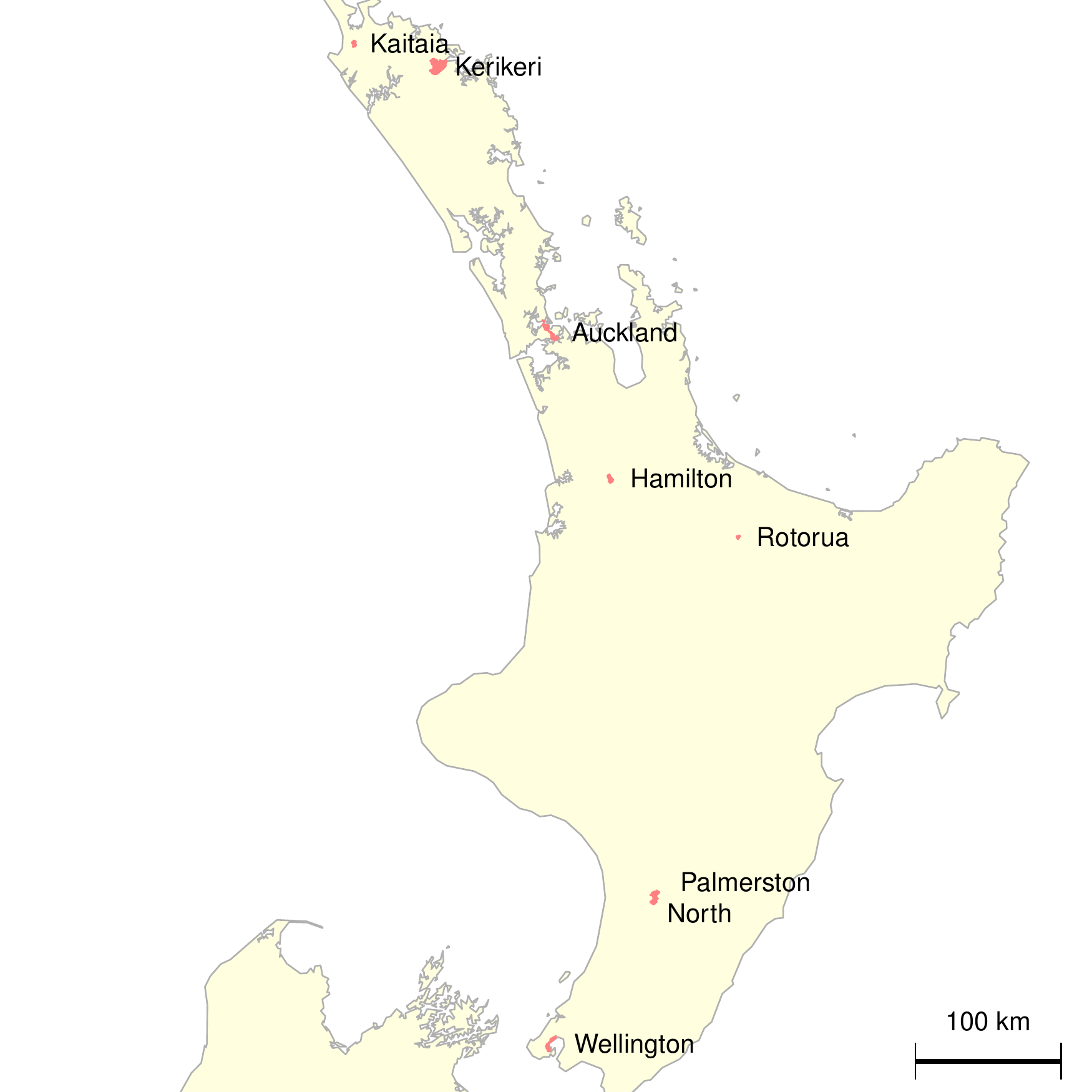}
\end{center}
\caption{Loci in the North Island strong component.
} 
\label{fig:loci-nz}
\end{figure*}

Figure~\cite{fig:loci-nz} is a spatial visualization of the loci in the largest strong component for New Zealand. Loci appear in central Auckland, Wellington, Palmerston North, Hamilton, Kerikeri, Kaitāia, and Rotorua. Local spatial representation of these loci is shown in Figure~\ref{fig:loci-maps} (Rotorua is not pictured because it is a singelton). Although the detection of loci in major cities is intuitively expected, even more rural towns like Kerikeri and Kaitāia with population of only around 8,000 and 6,300, respectively, are hosts of loci which is consistent with their role as hubs in a more sparsely populated region. The loci are places individuals are commuting \emph{to and through} in greater proportion compared to other areas. This may translate into these areas requiring greater investment in upkeep, since they are more heavily used, and it likely translates into proportionally greater aggregate delays when movement through these areas is hampered by issues like construction and congestion.

\section{Stationary distribution and loci as global graph properties}

The stationary distribution of the strong component as well as the loci, as described here, describe global properties of the mobility graph. Intuitively, this can be seen from Section \ref{calc_stat_dist} where the linear-optimization is over the probability transition matrix of the entire graph. The loci feature inherits the global property characterization since it is a subset of the stationary distributions, which is global. These properties, can also be seen as properties of individual, corresponding vertices: stationary distribution (proportion of times individuals go to and through that vertex) and loci status (yes or no).

The procedures for calculating both the stationary distribution of the mobility graph and the loci encode novel feature information in the mobility literature and are not represented by existing supervised or unsupervised procedures. As a result, a direct comparison between it and similar procedures is not possible. However, it remains important to evaluate these features in terms that allow us to distinguish them from other related approaches and provide a more intuitive understanding of the mobility information that they encode. To do this, we assess how much information is encoded in this feature compared to four standard {\em local} features that can be classified as either local properties of vertices or features that are calculated from the user trajectories, the sequence of SA2s defining a commute in this paper. The first local graph vertex property is the \emph{in-degree} of each vertex, i.e., the number of edges going from neighbors of an SA2 area into the SA2 area. The second is the \emph{weighted in-degree}, which multiplies each in-edge by the number of commuters coming into the SA2 area.  The first trajectory feature will be termed the \emph{total incoming traffic}. It counts the total number of commuters going \emph{to} (and not \emph{through}) an SA2 area  from any location. It is equivalent to the weighted in-degree of the source-destination network. The fourth and final is the \emph{total traffic} which measures how many commuters start at, go through, or end at a given SA2 area. This is calculated by taking all of the trajectories and counting how many include a given SA2. Together, these four features will be referred to as the {\em comparison features}.

This section provides three results supporting the claim that both the stationary distribution as well as the loci provide distinct and potentially valuable information when understanding population-scale human mobility. First, up to approximately one-quarter of the variation in the stationary distribution is captured by the trajectory features. This implies that the global features encoded in the stationary distribution are partially captured by information in individual trajectories. Second, because most of the variation cannot be explained in terms of comparison features, the global features capture information that is distinct from them. This information is likely from the graph structure, which cannot be recovered from the local graph or trajectory features presented. Third, when the trajectory features are regressed onto the stationary distribution, the loci have larger corresponding residual values thereby reinforcing the claim that loci identify vertices where there is more than expected mobility.


Admittedly, the fact that these four local features fail to encode the global features being proposed does not imply that there are not other features that do. There are not general theorems telling us under which conditions global properties can be recovered from the features we are comparing against. The claim that they cannot in this case, is left as a conjecture.

\begin {table}
\begin{center}
\begin{tabular}{rrrr} 
{\bf Metric} & {\bf p-value} & {\bf Linear adj. R$^2$} & {\bf RF OOB R$^2$.} \\ \hline
In-degree & 0.005 & 0.005 & 0.001 \\
Weighted in-degree & 0.561 & $\le 0.001$ & $\le 0.001$ \\
Total incoming & $<$ 0.001 & 0.143 & 0.150 \\
Total traffic &  $<$ 0.001 & 0.249 & $\le 0.001$ \\
\end{tabular}
\end{center}
\caption {Association between local graph features and stationary distribution. Linear adj. R$2$ is the adjusted R$^2$ value of the linear regression. RF OOB R$^2$ is the out-of-bag R$^2$ value of the random forest model.xx`} \label{tab:linear-model} 
\end{table}

\begin{figure*}
\begin{center}
\includegraphics[width=0.8 \textwidth]{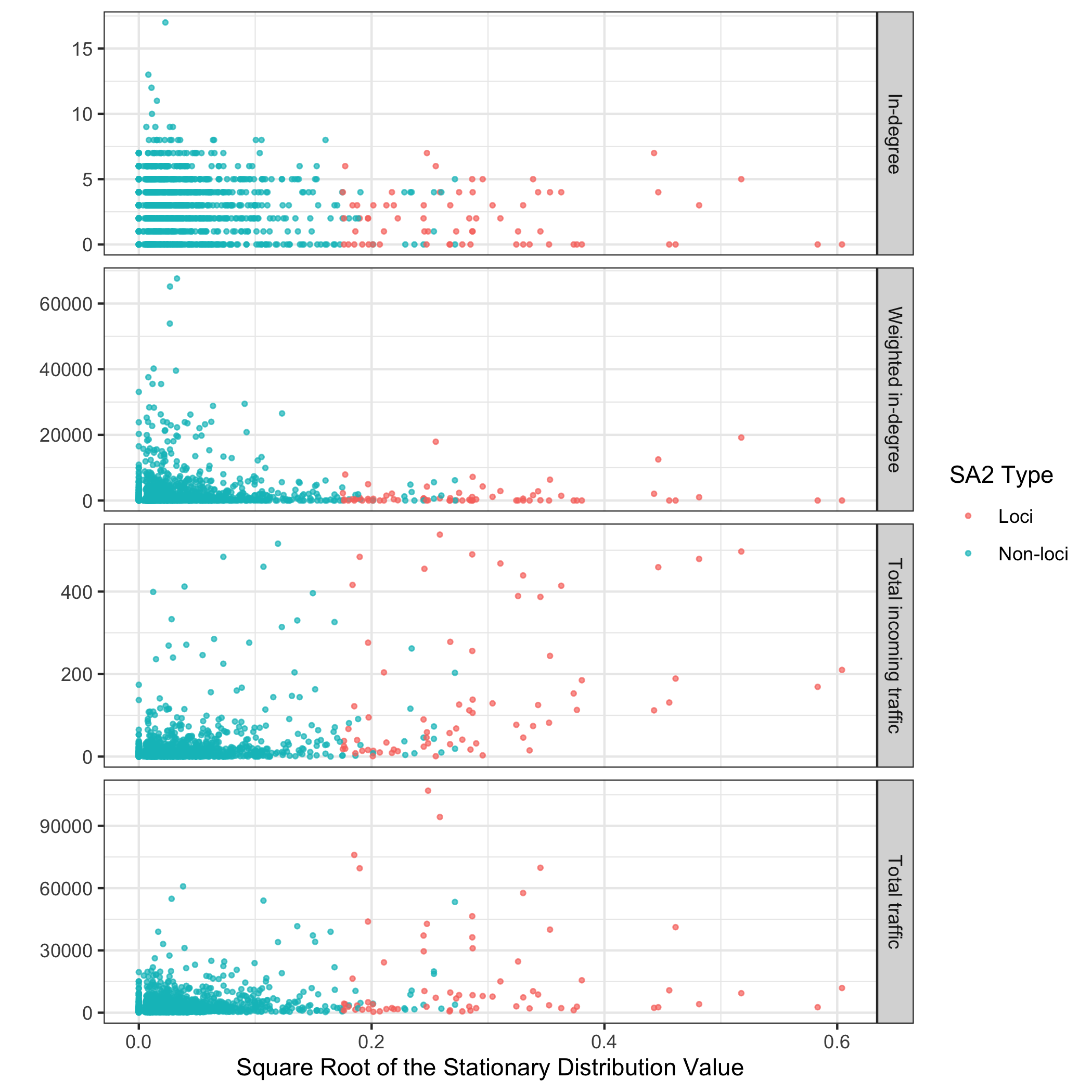}
\end{center}
\caption{Scatter plots of the local features vs. the square root of the stationary distribution. \label{fig:local-global-scatter}}
\end{figure*}

Figure~\ref{fig:local-global-scatter} shows a scatter plot of the the square root of the stationary distribution ($x$-axis) versus each of the comparison features ($y$-axis), along with the loci status of each SA2 area. The graph shows that loci tend to have higher stationary distribution values but not necessarily higher comparison-feature values. Table \ref{tab:linear-model} shows the associative relationships between each of the comparison features and the stationary distribution. The first column is the $p$-value of the slope coefficient of the linear model regressing the comparison feature onto the stationary distribution. The second column gives the adjusted R$^2$ values of the associated linear models. The third column gives the out-of-bag R$^2$ values of the associated random forest model. The third column is provided to show that the associative information exists on a linear subspace and can be well-characterized using a linear model. Other linear model regression diagnostics were performed, including a QQ-plot of the residuals, which showed that the residuals were approximately normal but did contain heavy, symmetric tails. The table shows that, except for the ``Weighted in-degree'' variable, each of the comparison features had significant linear association with the stationary distribution. However, the comparison features were not able to account for a large portion of the variation with the ``Total traffic'' variable accounting for 24.9\% and others accounting for less. 

\begin {table}
\begin{center}
\begin{tabular}{rrr} 
{\bf Metric} & {\bf Lin. adj. R$^2$} & {\bf RF OOB R$^2$} \\ \hline
Total inc. $\sim$ Total traffic & 0.334 & 0.193 \\
Stat.\,Dist. $\sim$ Total inc. + Total traffic & 0.261 & 0.230 \\
\end{tabular}
\end{center}
\caption {Association between comparison distribution and traffic.} \label{tab:title} 
\end{table}

Since Table \ref{tab:linear-model} shows that ``Total incoming'' and ``Total traffic'' are associated with the stationary distribution, similar linear and non-linear models were constructed to quantify the associative relationship between these variables. Table \ref{tab:title} shows the relationship between the two trajectory features (``Total incoming'' and ``Total traffic''); and, the relationship between the stationary distribution and the two trajectory features together. The ``Total incoming'' variable explains 24.9\% of the in-sample variation of the stationary distribution using a linear model and 14.4\% of the out-of-sample variation using a random forest model, while ``Total traffic'' variable explains 14.3\% and 0\% of the variation, respectively (Table \ref{tab:linear-model}). ``Total traffic'' accounts for 33.4\% and 19.3\% of the variation in ``Total incoming traffic'', and the combined trajectory features account for 26.1\% and 23.0\% of the variation in the stationary distribution, respectively (Table \ref{tab:title}). From this we can conclude that although some of the variation in the stationary distribution can be explained by the trajectory features, it is only about one quarter of the stationary distribution's total variation. 

\begin{figure*}[htbp]
\begin{center}
\includegraphics[width=0.8 \textwidth]{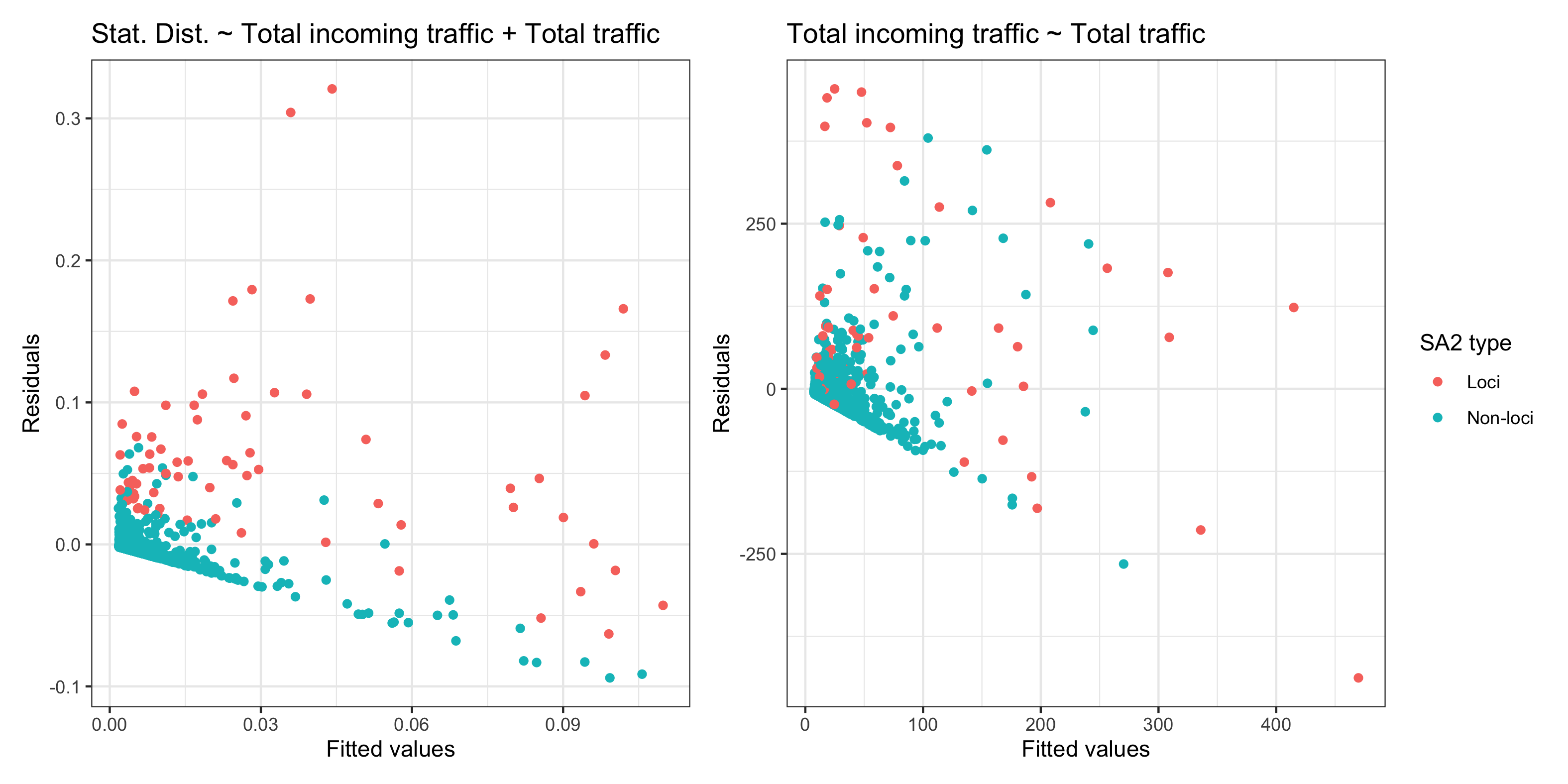}
\end{center}
\caption{Residual plots from regressing (a) stationary distribution on trajectory features, and (b) total incoming traffic on total traffic (trajectory features). \label{fig:local-global-residuals}}
\end{figure*}

Figure~\ref{fig:local-global-residuals} plots the fitted vs. residuals values for the combined linear model on the left and the regression between the trajectory features on the right. From the plot on the left we can see that the loci (colored in red) tend to have large, positive residual values. This is likely because the global structure, which is not captured by ``Total incoming traffic'' and ``Total traffic'' plays a large role in determining the vertices with large stationary distribution values, resulting in large, positive residuals. The plot on the right shows that large residuals between the trajectory feature model is not sufficient for identifying loci since they are interspersed with non-loci residuals.

\section{Conclusions}

As shown, mobility loci identify locations to which people move to and through at a rate higher than expected given the movement and structure of the underlying mobility graph. For the  New Zealand commuting data, we identified SA2 areas considered to be mobility loci. These corresponded to specific, high-population and high-movement areas in Auckland and Wellington as well as less populated areas of high mobility. For infrastructure planners, these could be seen as places with high ``commuting stress'' or places where, when they experience delays, have greater impact on commute time on the population than others. Information, similar to what was presented here, could then be incorporated into impact assessment with planners proposing the expansion of infrastructure based on whether or not an investment alleviates the associated commuting stress or removes the loci entirely. Additionally, the two vertex features presented here (stationary distribution and loci status) can be used to study area-level associations with applications in epidemiology and urban planning, among others.

As a final note, it should be pointed out that the underlying procedure for identifying loci is by no means limited to mobility graphs and should extend readily to other domains. For example, research in social networks tends to focus on community detection, encoded as an undirected graph with the goal of identifying members of a community. 
Other areas of application include transportation networks, genomic pathway analysis, or other areas whose representation may be a directed, potentially weighted graph.

\section{Software implementation}

All aspects of the analysis presented were implemented using the R Programming Environment \citep{rcore}. Data formatting and shaping relied on the \texttt{dplry} \citep{dplyr}, \texttt{tibble} \citep{tibble}, and \texttt{tidyr} \citep{tidyr} packages with variable checking was performed using the \texttt{checkmate} \citep{checkmate} package. The construction of routes between home and work SA2's was performed using the \texttt{ghroute} \citep{ghroute} package. Parallel processing, employed to speed-up the permutation procedure and other analyses, used the \texttt{foreach} \citep{foreach} package with the using the multicore backend provided by \texttt{doMC} \citep{doMC}. Plots were created using \texttt{ggplot2} \cite{ggplot2} and \texttt{patchwork}
\citep{patchwork}. Spatial visualizations were created using \texttt{proj4}
\citep{proj4}, \texttt{sf} \citep{sf}, and \texttt{snippets} \citep{snippets} with \texttt{RColorBrewer} \citep{RColorBrewer}
providing color mappings. Mobility graphs were represented and processed and analyzed using a combination of packages \texttt{igraph} \cite{igraph}, \texttt{tidygraph} \cite{tidygraph}, \texttt{Matrix} \citep{Matrix}, and \texttt{graphmobility} \citep{graphmobility}, the last of which encapsulates the novel analysis aspects of this paper. Finally, the \texttt{randomForestSRC} \citep{randomforestSRC} packages was used for the Random Forest analysis.

\begin{funding}

The first author was supported by the National Science Foundation (NSF) Grant Human Networks and Data Science - Infrastructure (HNDS-I), award numbers 2024335.

The second author was supported by the National Science Foundation (NSF) Grant Human Networks and Data Science - Infrastructure (HNDS-I), award numbers 2024233.

\end{funding}

\bibliographystyle{imsart-nameyear.bst} 
\bibliography{bibliography}       

\end{document}